%% file: WGL.tex
\documentclass[aps,prl,reprint,showpacs,floats,floatfix,amsmath,amssymb]{revtex4-2}%
\usepackage{hyperref}

\usepackage{graphicx}
\usepackage{physics}
\usepackage{xcolor}
\usepackage{bm}

\begin{document}
\author{R. Whelan} \affiliation{School of Physics and Astronomy, University
  of Birmingham, Edgbaston, Birmingham, B15 2TT, United Kingdom}
\author{D. M. Gangardt} \affiliation{School of Physics and Astronomy,
  University of Birmingham, Edgbaston, Birmingham, B15 2TT, United Kingdom}
\author{I. V. Lerner} \affiliation{School of Physics and Astronomy, University
  of Birmingham, Edgbaston, Birmingham, B15 2TT, United Kingdom}
\title{Majorana-like Josephson branch structure in one-dimensional transport of cold atoms}
\begin{abstract}
We study phase-controlled transport through a one-dimensional channel of attractively interacting fermions connecting two three-dimensional BCS reservoirs. The competition between channel phase stiffness and pair tunnelling at the two contacts results in the emergence of two  $4\pi$-periodic families of stationary states, characterised by symmetric or asymmetric phase drops at the boundaries.  Advancing the reservoir phase difference by $2\pi$ exchanges these families so that the equilibrium energy remains $2\pi$ periodic, while the $4\pi $ periodicity might be attained in a diabatic regime. This structure admits a formal description in terms of two boundary Majorana operators, combining into a nonlocal fermion mode. Renormalization-group analysis shows that quantum fluctuations and phase slips are suppressed in the zero-temperature thermodynamic limit, preserving the branch crossings. The predicted branch exchange, metastable currents, and sweep-rate-dependent switching provide experimentally testable signatures of Majorana-like Josephson branch physics in a gapless interacting channel.

\end{abstract}

\maketitle
\newcommand{\e}{\mathrm {e}}

One-dimensional (1D) quantum systems provide a setting in which interactions qualitatively change the nature of both equilibrium states and transport. Instead of the quasiparticles characteristic of a Fermi liquid, their low-energy excitations are collective density and phase modes described by Tomonaga--Luttinger liquid theory \cite{Giamarchi2003}. In particular, connecting macroscopic superconducting reservoirs through such a 1D channel has long been known to produce transport properties very different from those of a conventional Josephson junction.

For an interacting electronic wire coupled to two superconductors, the Josephson current is strongly affected by the Luttinger-liquid correlations and by the nature of the contacts \cite{Fazio1995,Maslov1996}.
  In such a case, perfect Andreev reflection imposes boundary conditions which pin the bosonic phase field to the phases of the adjoining superconductors. The corresponding problem is naturally formulated as a boundary sine-Gordon theory, resulting in a nonanalytic dependence of the critical Josephson current  \cite{Fazio1995,Maslov1996}.

\begin{figure}[!t]
\noindent\includegraphics[width=\columnwidth]{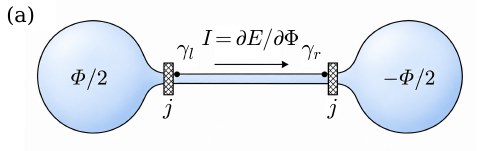}\\[-3pt]
\includegraphics[width=\columnwidth]{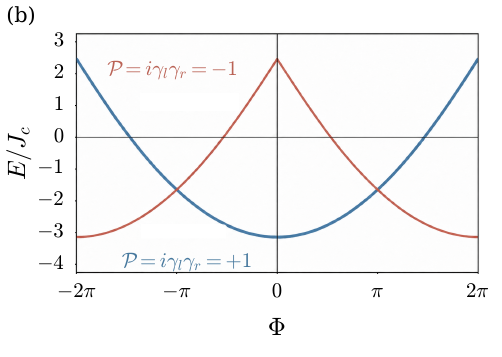}
\caption{
(a) A one-dimensional channel of attractively interacting fermions
connects two three-dimensional BCS reservoirs with phases
$\Phi/2$ and $-\Phi/2$. The junctions have tunnelling amplitude $j$,
and the contact labels $\gamma_l$ and $\gamma_r$ refer to the
formal Majorana representation of the branch label discussed in the text.
The arrow indicates the stationary pair superflow,
$I=\partial E/\partial\Phi$ (in units with $\hbar=1$).
(b) Josephson energy branches labelled by
$\mathcal{P}_+ =+1$ (blue, S) and ${\mathcal{P}}_-=-1$ (red, A),
shown for $W_0=\pi/2$; see Eqs.~(\ref{E_kin}) and
(\ref{even-odd}).
The family-resolved energy structure is $4\pi$-periodic, whereas
the equilibrium lower envelope is $2\pi$-periodic.
} \label{Epng}
\end{figure}

  Ultracold atomic gases offer an especially clean realization of such physics. Mesoscopic transport geometries in which two macroscopic reservoirs of
ultracold fermions are connected by a narrow channel, quantum wire, or
quantum point contact have been realized experimentally
\cite{Brantut:2012,Brantut:2013,Krinner2015,Valtolina2015,Husmann2015,%
Krinner2016,Hausler2017,Burchianti2018,Luick2020}.
In particular,  experiments with
strongly interacting Fermi superfluids have demonstrated transport dominated
by pairing and Andreev processes \cite{Husmann2015,Krinner2016}.

In such experiments,
interaction strength, geometry, and reservoir populations can all be controlled independently. Also, transport experiments in one-dimensional attractive Fermi gases have provided evidence \cite{Lebrat2018} for the Luther--Emery regime \cite{LutherEmery1974} with gapped spin excitations and a gapless   particle-number mode. In this regime enhanced singlet-pairing correlations result in a state which is the one-dimensional analogue of a superconductor, although true long-range order is destroyed by phase fluctuations. These developments make it possible to explore superconducting boundary phenomena and strongly correlated one-dimensional transport in a regime that is difficult to access in electronic systems.

In this work we consider attractive spin-$1/2$ fermions with Fermi momentum $k_{\mathrm{F}}$ propagating through a 1D channel of length $L$ connecting two 3D reservoirs containing fermionic superfluids in the BCS state.  We assume that  the stationary flow is driven entirely by an imposed phase difference  $\Phi $  between the reservoirs \footnote{Phase imprinting provides a route to preparing a phase bias \cite{Luick2020}. Population imbalance can also generate phase evolution \cite{Valtolina2015}; a stationary phase bias requires the reservoir chemical potentials to be equal.}, as illustrated  in Fig.~\ref{Epng}(a). We show that in contrast to the repulsive case considered in Refs.~\cite{Fazio1995,Maslov1996},  the physics is governed by
the nontrivial saddle-point configuration, which is determined by
 the competition between the stiffness of the 1D liquid and the boundary tunnelling amplitudes. The boundary renormalization group (RG) flow favors phase pinning at low energies, confirming robustness of this stationary-state structure.

The  manifestation of this physics is a $4\pi$-periodic Josephson response, analogous to systems containing spatially separated Majorana modes  \cite{Kitaev2001,FuKane2009}. This resemblance  can be ascribed to the emergence of the nonlocal fermionic mode formed by a pair of local boundary Majorana operators $\gamma _{l,r}$ at the left or right contact \footnote{Although such an analogy  is useful, a $4\pi$-periodic
Josephson energy structure can also arise without Majorana zero
modes \cite{ChiuDasSarma2019}.}  , similarly  to the case of
a topological superconducting junction \cite{Lutchyn:2010}. The situation considered here is different in a {{crucial}} respect: the one-dimensional bulk is not a fully gapped mean-field topological superconductor. Its spin excitations are gapped, but its number sector remains a gapless Luttinger liquid \footnote{This is similar to the 1D model of spinless fermions with attraction considered in Ref.~\cite{Debortoli2026}.}.

The central result is the emergence of a  two-family Josephson energy structure, as illustrated in Fig.~\ref{Epng}(b). In the symmetric (S) family, the boundary  phase drops are equal on both contacts, while in the asymmetric one they differ by $2\pi $.
 The Majorana operators form the nonlocal parity operator
 \begin{align}\label{P}
\mathcal{P}=i\gamma_l\gamma_r \,,
\end{align}
with eigenvalues ${\mathcal{P_\pm}}=\pm1$ distinguishing these  two families which are interchanged by a  $2\pi$ phase shift.
In the thermodynamic limit, the nonlocal parity conservation prevents  these  branches from hybridizing at their crossing, resulting in  quasidegenerate ground states for $\Phi =\pm\pi $ and producing the fractional Josephson effect.

\paragraph{Model and effective boundary action.} The setup outlined above (Fig.~\ref{Epng}(a)) is described by the   action
\begin{align}\label{Stotal}
    S=S_l+S_r+S_{\mathrm{tun}}+S_{\mathrm{LL}}\,.
\end{align}
Here $S_{\mathrm{LL}}$ is the Luttinger-liquid action in the channel, $S_{\mathrm{tun}}$ is the tunnelling action, and $S_{l,r}$ are the standard BCS actions in the reservoirs:
\begin{align}\label{BCS}
S_{\lambda}
=
-\int d\xi _1\, d\xi _2\;
\overline{\Psi}_{\lambda}(\xi _1)\,
 {G}^{-1}_{\lambda}(\xi _1,\xi _2)\,
\Psi_{\lambda}(\xi _2).
\end{align}
Here $\lambda =l,r$ labels the left and right reservoirs,  $\xi \equiv (\bm r, \tau)$, $\int d\xi \equiv \int_0^\beta d\tau \int d\bm r$ ($\tau$ is imaginary time, $\beta =1/k_{\mathrm{B}}T$ is inverse temperature) and the Nambu spinors and  BCS kernel are
\begin{align*}
\!\!\! \Psi_\lambda {=}
\begin{pmatrix}
\Psi_{\lambda\uparrow} \\
\overline{\Psi}_{\lambda\downarrow}
\end{pmatrix}\!,\;
\overline{\Psi}_{\lambda}{=}\left(
\overline{\Psi}_{\lambda \uparrow},
\Psi_{\lambda \downarrow}
\right)\!,\;
 {G}^{-1}_{\lambda}
{=}
\begin{pmatrix}
-{\partial}_{+} & \Delta_{\lambda} \\
\Delta_{\lambda}^{*}  &-{\partial}_{-}
\end{pmatrix},
\end{align*}
where  $\partial_\pm\equiv \partial_\tau\pm\qty(\nabla^2/2m+\mu )$. We assume in what follows that $\Delta_\lambda =|\Delta|\, e^{i\phi_\lambda } $, i.e.\ the magnitude of the order parameter, as well as the chemical potential $\mu $ and the normal density of states $\nu _0$, are the same in both reservoirs,  implying that the current through the 1D channel is driven by a constant phase difference  $\phi _l-\phi _r \equiv \Phi$. We choose,  without loss of generality,  $\phi _{l}= -\phi _r=\Phi /2$.

The reservoir fields are coupled to the fermionic fields in the 1D channel,  $
 {{\psi}} (x, \tau)  =
\left(
{\psi}_{  \uparrow}(x, \tau),
\overline{\psi}_{  \downarrow}(x, \tau)
\right)^{\mathrm T}
$ by the   tunnelling  action
\begin{align}\label{Stun}
    S_\text{tun} &=-j\int  _0^\beta \!    \dd\tau\sum_{\lambda=l,r }   \left[ {\overline\Psi}_{\lambda }( \tau) \psi_{\lambda }(\tau) +\text{ h.c.}\right].
\end{align}
Here $ {\overline\Psi}_{\lambda }( \tau)$ and $ {\psi}_{\lambda}\qty(\tau)$  are the reservoir and channel    fields  evaluated at the reservoir contact point $\bm r_\lambda $
 corresponding to the channel endpoint $x_\lambda =\pm L/2$. The tunnelling amplitudes, $j$, are assumed to be spin-independent  and equal on the left and right barriers.

Performing Gaussian integration over the reservoir fields reduces the first three terms in Eq.~\eqref{Stotal} to the  effective boundary action:
\begin{align}
S_{\mathrm{b}}
&= -\tfrac12
j^2
\sum_{\lambda=l,r}
\int  \dd\tau\, \dd\tau'\,
 {\overline{\psi}}_{\lambda}(\tau)\,
 {G}_{\lambda}(\tau-\tau')\,
 {\psi}_{\lambda}(\tau').\label{eff}
\end{align}
Here $ {G}_{\lambda}=\begin{pmatrix}
\mathcal{G}_{\lambda} & \mathcal{F}_{\lambda} \\
\mathcal{F}_{\lambda}^{\dagger} & -\mathcal{G}_{\lambda}
\end{pmatrix}$ is the Nambu Green's function of lead $\lambda$. Only  anomalous propagators,
$\mathcal{F}_{l,r}$,
carry the superconducting phases.
For a bulk BCS lead, the anomalous propagators have the Matsubara-frequency representation, reduced in the low-energy limit $|\omega_n|,T\ll\Delta$,  to
\begin{align}\label{BSCGF}
\mathcal{F}_{l,r}(\omega_n)
=-\pi\nu_0
\frac{\Delta\,e^{\pm i\Phi/2}}
{\sqrt{\omega_n^2+\Delta^2}}
\approx
-\pi\nu_{0}
e^{\pm i\Phi/2},
\end{align}
where $\nu_0$ is the  normal-state density of  states in the reservoirs. Then
 $\mathcal{F}_{l,r} (\tau-\tau')\approx- \pi \nu _0  e^{\pm i\Phi /2 } \delta  (\tau-\tau')$ and the phase-dependent part of the boundary action (\ref{eff})  becomes local in imaginary time:
\begin{align}\label{Slr}
\!S_{{\mathrm{b}}}
\simeq
 \frac{\pi\nu_0 j^2}2\sum_{\lambda =l,r}
\int \nolimits_0^\beta \! \! \dd\tau  \,
\Big[
e^{- i\phi _\lambda }
\psi_{\lambda \downarrow}(\tau)\psi_{\lambda \uparrow}(\tau)
+
\mathrm{h.c.}
\Big].
\end{align}

We assume from the outset that attractively interacting fermions in the 1D channel
are in the Luther--Emery regime where the backscattering interaction
opens a spin gap $\Delta_\sigma$.  In this case, as we show in the \emph{End Matter}, Eq.~\eqref{Slr} reduces  after bosonizing the channel fields $\psi$ to
\begin{align}
    S_{\mathrm{b}}&=  -J_0
    \int\nolimits_0^\beta \!\!\dd\tau\left[\cos\chi  _l(\tau)+\cos\chi  _r(\tau) \right],\label{ABC}
\end{align}
where   $J_0\sim  \nu _0k_{\mathrm{F}} j^2$
 is the effective tunnelling energy.  The boundary phase drops $\chi _\lambda \qty(\tau)$ are related to the boundary values   $\varphi _\rho^{\lambda } $  of the      bosonic phase field  $\varphi _{\rho }\qty(x, \tau)$  in the number ($\rho$)  sector by
\begin{align}\label{phasedrops}
\chi_l(\tau)&=\tfrac12\Phi -2\varphi_\rho^l(\tau),
&\chi_r(\tau)&=\tfrac12\Phi+2\varphi_\rho^r(\tau) .
\end{align}
  In  the low-energy Andreev strong-coupling limit, the boundary dynamics can  be expressed solely in terms of the number-phase field $\varphi _\rho$, as shown in the \emph{End Matter}.  The bosonized Luttinger-liquid action entering Eq.~\eqref{Stotal} then reduces to
\begin{align}\label{LL-eff}
    S_{\mathrm{LL}}[{\varphi _\rho}]=\frac{K }{\pi v } \int\limits_0^\beta \dd \tau \!\!\!\int \limits _{- L/2}^{ L/2} \!\!\! \dd x\left[\qty(\partial_\tau\varphi _\rho)^2+v ^2\qty(\partial_x\varphi _\rho)^2\right].
\end{align}
Here $K$ and $v$ are the  Luttinger parameter and (renormalized) Fermi velocity  in the channel.
Equations (\ref{ABC})--(\ref{LL-eff}), which govern transport of cold fermions with attraction, correspond to those in Ref.~\cite{Simpson2014} for transport of cold bosons with repulsion.

\paragraph{Saddle-point configuration.} For the attractive case considered here, $K>1$, perturbative approach in weak tunnelling, used for repulsive interactions  in Refs.~\cite{Maslov1996,Fazio1995}, is not applicable in the limit $L\to\infty$, thus motivating a nonperturbative saddle-point treatment.

A  static saddle point is found by varying the bulk action, Eq.~\eqref{LL-eff}, which gives
$
\partial_x^2\varphi_\rho(x)=0\,.
$
Hence the phase profile is  linear along the channel. With the boundary conditions (\ref{phasedrops}), we find the saddle-point solution as
\begin{align}
\varphi_{\rho0}(x)&=
\frac{(\chi_+-\frac{1}{2}\Phi )x}{L}+\frac {\chi _-}2, &\chi _\pm\equiv \frac {\chi _r\pm\chi _l}2,
\label{saddle}
\end{align}
describing uniform superflow through the one-dimensional channel.

Substituting the saddle-point configuration (\ref{saddle}) into    Eqs.~(\ref{ABC}) and (\ref{LL-eff}), we obtain the   energy
\begin{align}\label{E_kin}
E/J_{\mathrm{c}}&= \qty(\tfrac12\Phi-\chi_+)^2-
2 W_0  \cos\chi_+\cos\chi_-,
\end{align}
where $J_{\mathrm{c}}\equiv J_{\mathrm{c}}\qty(L)=  {K  v }/({ \pi L})$ is the phase stiffness of a channel of length $L$, and
$ W_0
\equiv
 {J_0}/J_c$ is the dimensionless tunnelling strength. The first term in Eq.~\eqref{E_kin} is the energy of superfluid flow through the channel and the second is an effective two-parameter washboard potential.
Minimizing $E$ with respect to  $\chi_+$ and $\chi_-$ gives
\begin{align}\notag
 W_0\sin\chi_+\cos\chi_-&=
\tfrac12\Phi-\chi_+
,
 \\\label{cos}
\cos\chi_+\sin\chi_-
&=
0.
\end{align}
The locally stable minima have $\chi_-=0$ or $\pi$ modulo $2\pi$, with $\cos\chi_+\cos\chi_->0$ \footnote{The solutions with $\cos\chi_+=0$ are  saddle points rather than minima}. This gives two stationary-state families:
\begin{align}\label{even-odd}
\tfrac12\Phi-\chi_+^{{\mathrm{S,A}}}
&=\left\{\begin{matrix}
           + W_0\sin\chi_+^{{\mathrm{S}}} , &\chi_-=0,  \\
         - W_0\sin\chi_+^{\mathrm{A}},& \chi_-=\pi, \\
         \end{matrix}\right.
\end{align}
corresponding, respectively, to symmetric ($S$) and asymmetric ($A$)
phase drops at the two boundaries.  In the symmetric family
$\chi_r=\chi_l$, while in the asymmetric family
$\chi_r=\chi_l+2\pi$.

\begin{figure}
    \includegraphics[width=\columnwidth]{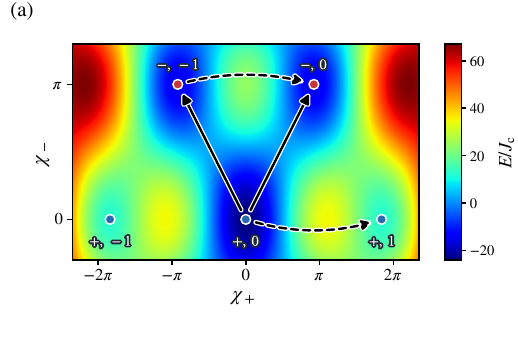}\\[-3pt]
\includegraphics[width=\columnwidth]{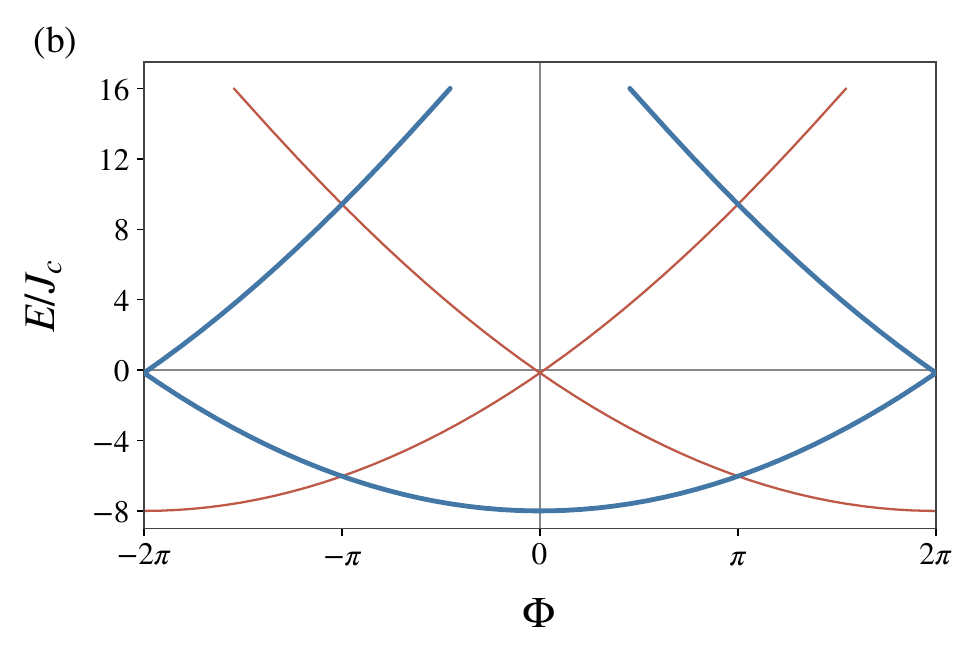}
  \caption{(a) Colour density plot of the washboard energy in
Eq.~\eqref{E_kin} at $\Phi=0$ and $W_0=12$.
 Dots mark locally stable minima,
labelled $({\mathcal{P}},m)$: $n=2m$ for ${\mathcal{P}}_+=+1$ (blue dots, S)
and $n=2m+1$ for ${\mathcal{P}}_-=-1$ (red dots, A).
Solid arrows connect opposite winding-parity families
($\Delta n=\pm1$, one-contact phase slips); dashed arrows connect
different winding states within the same family
($\Delta n=2$, $\Delta m=1$, phase slips at both contacts).
(b) Saddle-point energies for $W_0=4$. In this strong-coupling regime,
each family contains several metastable branches distinguished
by the integer winding number $m$.
}\label{4a}
\end{figure}

An important consequence of   Eq.~(\ref{even-odd}) is that the two stationary-state families have energies  which are intrinsically $4\pi$-periodic in the  reservoir phase difference $\Phi$:
 a shift
$\Phi\rightarrow\Phi+2\pi$ does not return a given family to itself;
instead, it interchanges the symmetric and asymmetric families, $
E_{\mathrm{{S,A}}}(\Phi+2\pi)=E_{\mathrm{{A,S}}}(\Phi).
$
A further shift by $2\pi$ returns to the original family,
$
E_{\mathrm{{S,A}}}(\Phi+4\pi)=E_{\mathrm{{S,A}}}(\Phi),
$
see Figs.~\ref{Epng}(b) and \ref{4a}(b), where we  draw the  branches only in the regions where solutions are (meta)stable.

\paragraph{Winding families and Majorana representation.}
 The branch label admits a formal representation using two self-adjoint Majorana operators $\gamma_l$ and $\gamma_r$, with
$\{\gamma_\lambda,\gamma_{\lambda'}\}=2\delta_{\lambda\lambda'}$. Introducing nonlocal fermion operators
\begin{align}\label{f}
f=\frac{\gamma_l+i\gamma_r}{2},\qquad
f^\dagger=\frac{\gamma_l-i\gamma_r}{2},
\end{align}
we express the parity operator of Eq.~\eqref{P} as $ {\mathcal{P}}=2f^\dagger f-1$.
For sufficiently large dimensionless tunnelling strength $ W$,  each of the two families, labelled by eigenvalues  $\pm1$ of operator $\mathcal{P}$, contains several metastable minima as in Fig.\ref{4a}, corresponding to the phase winding,  $\chi _+\to \chi _++2\pi m$, where    $m$ is
  an integer winding index.

Avoided crossings between different branches could be generated by phase-slip processes (instantons) connecting different  minima. Equivalently, we can consider fluctuations  beyond the saddle-point approximation via the RG approach.

We  show  in the \emph{End Matter} that such fluctuations are strongly suppressed in the thermodynamic limit thus restricting the scope for avoided crossings. The boundary RG flow for $K>1$ increases the dimensionless cosine term toward strong coupling. In its weak-coupling scaling regime,
$W\sim W_0 (L/a)^{1-1/K}$, with $a $ being an ultraviolet cutoff length \footnote{For repulsive interactions, $K<1$, the boundary coupling instead flows toward zero, consistent with a perturbative treatment \cite{Fazio1995,Maslov1996}.}. This flow favors phase pinning. We show that together with the strong-coupling phase-slip picture \cite{Schecter2012}, it supports the persistence of metastability and suppressed switching at low energies.

At finite size, phase-slip-induced mixing can produce avoided crossings between winding states. Slow evolution through such a crossing follows the hybridized state, whereas sufficiently rapid evolution can retain the metastable branch, as in Landau--Zener dynamics \cite{Dominguez2012,Pikulin2012}.   This might provide
a direct probe of the emergent Majorana physics.

The critical Josephson current behaves differently in such diabatic and adiabatic regimes, being of the order of $J_0(a/L)^{1/K} $ in the former, and $Kv/L$ in the latter, as demonstrated in the \emph{End Matter}.  In both regimes it is parametrically larger than the critical perturbative current proportional to $J_0^2/L$.

There is a useful comparison with the number-conserving spinless-fermion model of Ref.~\cite{Debortoli2026}, where Majorana-like edge physics is diagnosed through parity-dependent edge-to-edge single-particle correlations.  Identifying whether the winding-family label has a direct physical fermionic interpretation remains an open question.

Our results motivate phase-controlled transport experiments that probe metastable winding states, current reversals, and switching as a function of sweep rate, channel length, and temperature. These measurements would test the Majorana-like Josephson branch structure.

 \begin{acknowledgments}
 We are grateful to Boris Altshuler and Ivan Khaymovich for useful discussion. This work was supported by the Leverhulme Trust via grant No. RPG-2024-401.
 \end{acknowledgments}
 \input{WGL.bbl}

\section*{End Matter}
 \paragraph*{Effective action.} The  fermionic   fields in the 1D channel, $
 {\overline{\psi}}(x,\,\tau)  =
\left(
\overline{\psi}_{\uparrow},
\psi_{\downarrow}
\right)
$ and their Hermitian conjugate, are bosonised in the usual way  and expressed via   phase fields, $\varphi _{\rho,\sigma}$, and canonically conjugate displacement fields, $\vartheta_{\rho,\sigma}$, in the number ($\rho$) and spin ($\sigma$) sectors as
\begin{align} \notag
\psi_{s}(x, \tau) &= \sqrt{\frac{k_{\mathrm{F}}}\pi } \!\!\sum_{\eta=\pm1}\!\!e^{ i (\eta k_{\rm F} - \frac\pi L)x} e^{- i( \eta\vartheta_\rho - \varphi_\rho  )}e^{- is( \eta\vartheta_\sigma - \varphi_\sigma )}
\end{align}
with $s{=}\pm1$ for spin $\uparrow,\,\downarrow$, $\eta{=}\pm1$ for right- and left-movers. Klein factors and particle-number zero modes are suppressed as they have no impact on this phase-only treatment of the pair-tunnelling action.

The attractive spin-$1/2$ channel
is assumed to be in the Luther--Emery regime, with a spin gap
$\Delta_\sigma$ and a gapless number sector~\cite{LutherEmery1974}. The slowly varying
singlet-pair operator is proportional to $
 e^{2i\phi_\rho}\cos(2\vartheta_\sigma).
$
Pinning of the spin displacement field $\vartheta_\sigma$ replaces its
contribution by a nonzero ground-state expectation value.
Its magnitude, together with short-distance normalization factors, is
absorbed into the contact pair-tunnelling energy.
The remaining number-sector action is
 \begin{align}\notag
S_{\rm LL}
=
\frac{1}{\pi}
\int \nolimits_0^\beta \!  \dd\tau \!\!\!\int\nolimits_{-L/2}^{L/2}\!\! \dd x
&\Big[
-2i\,\partial_x\varphi_\rho\,\partial_\tau\vartheta_\rho
+
v  K (\partial_x\varphi_\rho)^2
\\&+
\frac{v }{K }(\partial_x\vartheta_\rho)^2
\Big],\label{LE}
\end{align}
where $K$ and $v$ are the  Luttinger parameter and (renormalized) Fermi velocity  in the channel.

Gaussian elimination of the displacement-field fluctuations in
Eq.~\eqref{LE}  gives the phase-only action of Eqs.~(\ref{ABC}) and \eqref{LL-eff} with the effective boundary conditions of Eq.~\eqref{phasedrops}.

\paragraph*{Suppression of fluctuations.}
Consider fluctuations of the boundary  and  channel fields around the saddle-point solution (\ref{saddle}). Integrating out the fluctuating channel fields  in  the Gaussian phase action \eqref{LL-eff} yields
in the limit $
|\omega|
\gg
\omega_0\equiv v/L$  the Ohmic (Caldeira--Leggett) action~\cite{CL:81}
\begin{align}
S_{\mathrm{fl}}
= {K }
\int\frac{|\omega|\dd\omega}{(2\pi)^2}
\left[
|\delta\chi_+(\omega)|^2
+
|\delta\chi_-(\omega)|^2
\right],
\label{Ohmic}
\end{align}
  in terms of the fluctuating boundary fields,  $
\delta\chi_\pm(\tau)=\chi_\pm(\tau)-\chi _\pm^{\mathrm{S,A}} $.
 Their  fluctuations are governed  by the sum of the Ohmic action and of the following effective boundary action,
\begin{align}\label{Seff0}
S_{\mathrm{eff}}=J_{\mathrm{c}}\!\int_0^\beta \!\!\dd\tau \qty{
 \qty(\tfrac12\Phi{-}\chi_+)^2-
2 W  \cos\chi_+\cos\chi_-},
\end{align}
 obtained by integrating energy of Eq.~\eqref{E_kin}  over the imaginary time.
The nonlinear term in Eq.~\eqref{Seff0}  can be treated within the standard RG analysis, with the dimensionless tunnelling $W$ being the running parameter.

As only the boundary modes enter the action   (\ref{Ohmic})--(\ref{Seff0}), the bulk Luttinger parameter $K$ is not renormalized. The leading weak-coupling RG equation for the running dimensionless boundary coupling $W$ is obtained similarly  to the Kane--Fisher healing of a weak link for \(K>1\) \cite{KaneFisher1992PRL,KaneFisher1992PRB}:
\begin{align}
\frac{\dd\ln W}{\dd\ln b}
=1-\frac{1}{K },
\label{RG}
\end{align}
where $b$ is the length rescaling factor. In the weak-coupling scaling regime, this leads to
\begin{align}\label{W(L)}
W(L)\sim W_0(a) \qty(\frac La)^{1-1/K},
\end{align}
where  $W_0\qty(a) =J_0/J_{\mathrm{c}}(a)\equiv \pi  aJ_0/Kv $  with     $a$ being the ultraviolet cutoff.
This RG flow is analogous to dimensionless boundary tunnelling growth toward strong coupling, indicating a flow toward the Andreev strong-coupling regime discussed in Ref.~\cite{AffleckCauxZ}. The crossover length, $L_c$, to the strong coupling is estimated from $W\qty(L_c)\sim1$ yielding
\begin{align}\label{Lc}
L_c\sim aW_0(a)^{- K/(K-1)}\sim a\qty[\frac{Kv}{\pi a J_0}]^{  K/(K-1)}.
\end{align}

In the strong-coupling regime, we start with the pinned phase drops  parameterized by the winding index $n$ as $\chi _+=\pi n$ and $\chi _-=0,\,\pi $. The  leading winding energies are
\begin{align}
E_n^{(0)}(\Phi)
&=-2J_0
 +J_c\left(\frac{\Phi}{2}-\pi n\right)^2,
\label{em:winding}
\end{align}
which holds for fixed winding and phase bias
when $L\gg L_{\mathrm{c}}$, with the phase gradients still in the
low-energy regime.

An elementary $2\pi$ slip at the left or right contact changes
$(\chi_+,\chi_-)$ by $(\pi,-\pi)$ or $(\pi,\pi)$, respectively,
with the opposite signs for an anti-slip. Both processes change
$n$ by one. Substituting into Eq.~\eqref{Ohmic} a slip--anti-slip pair
at one contact $\lambda=l,r$, separated by imaginary time $\tau$ gives for $\tau_c\ll|\tau|\ll L/v$
\begin{align}
S_{\mathrm{pair}}(\tau)
&=2S_{\mathrm{core}}
 +2K\ln\!\left(\frac{|\tau|}{\tau_c}\right)\,.
\label{em:instanton}
\end{align}
Here $\tau_c$ is the short-time cutoff of the strong-coupling
description, and $S_{\mathrm{core}}$ is the  core
action for one slip. Equation~\eqref{Ohmic} identifies the phase-slip
operator dimension as $K$. Its dimensionless fugacity is
\begin{align}
\frac{\dd y_\lambda}{\dd\ln b}
&=(1-K)y_\lambda \; \Rightarrow\;
y_\lambda(L)\sim c_\lambda
 \left(\frac{L_{\mathrm{c}}}{L}\right)^{\!K-1},
\label{dualRG}
\end{align}
where $c_\lambda$ are  matching constants. This  dilute-slip
expansion for  $y_\lambda(E)\lesssim 1$ shows that the pinned
fixed point is stable for $K>1$.

The two limiting descriptions are
related in the same manner as weak tunnelling and weak
backscattering in the Kane--Fisher problem~\cite{{KaneFisher1992PRL,KaneFisher1992PRB}}, with
the strong-coupling endpoint corresponding to phase pinning and
Andreev reflection~\cite{AffleckCauxZ}.

\paragraph*{Critical Josephson current.}
The phase-controlled pair-number current at $T=0$ is
\begin{equation}\label{I}
I=\frac{\partial E}{\partial\Phi}
 =   \left(\frac{\Phi}{2}-\chi_+\right){J_{\mathrm{c}}}(L)
 =\pm J(L)\sin\chi_+\,,
\end{equation}
where $J_c(L)=Kv/(\pi L)$ and $J(L)=J_c(L)W(L)\sim J_0(a/L)^{1/K}$,
as   follows from Eqs.~(\ref{even-odd}) with the renormalized value of dimensionless tunnelling $W(L)$, Eq.~\eqref{W(L)}.

There are two limiting current protocols to be  distinguished. In a diabatic sweep through a small avoided crossing, the system can retain its winding branch rather than follow the hybridized equilibrium state. Equation~\eqref{I} bounds the magnitude of the stationary current by $J(L)$. If the chosen branch can be followed up to $|\sin\chi_+|=1$, its limiting current is
\begin{align}\label{Id}
I_{\mathrm{d}}=J(L) \sim J_0(a/L)^{1/K}.
\end{align}
For unequal tunnelling  energies $J_l$ and $J_r$, the stationary current must satisfy $I=J_l\sin\chi_l=J_r\sin\chi_r$, so $|I|\leq\min(J_l,J_r)$, i.e.\ the diabatic current is limited by the weaker contact.

 In an adiabatic limit, finite mixing or relaxation permits the system to follow the equilibrium lower-energy envelope, the maximum current occurs at the first branch crossing, $\Phi=\pm\pi$. For large renormalized $W$, Eq.~\eqref{even-odd} gives $\chi_+\simeq\pi/(2W)$ on the symmetric branch approaching $\Phi=\pi$, yielding
\begin{equation}\label{Ia}
I_{\mathrm{a}}\simeq\frac{\pi J_c}{2}=\frac{Kv}{2L}.
\end{equation}

 For both protocols, the limiting currents are parametrically larger than the critical perturbative current for repulsive interactions,
 $I_{\mathrm{crit}}\sim (J_0^2/\varepsilon _{\mathrm{F}}^2)(v/L)(a/L)^{2(1/K-1)}$ found in Refs.~\cite{Fazio1995,Maslov1996}.
 In contrast to the perturbative result, neither the diabatic nor the adiabatic critical current goes over to the noninteracting limit by simply taking $K\to1$ within the strong-coupling approximation. The reason for this is that the limits $K\to1$ and $L\to\infty$ do not commute.

Indeed, for $K>1$ the coupling is relevant and reaches the Andreev
strong-coupling regime at the crossover length $L_{\mathrm{c}}$, Eq.~\eqref{Lc}.
Hence, at any fixed $K>1$, taking $L\to\infty$
drives the junction to the Andreev fixed point, where the adiabatic envelope current
is controlled by the phase stiffness, $J_{\mathrm{c}}$, Eq.~\eqref{Ia}.
For weak bare tunnelling, $W_0(a)<1$, at fixed $L$ the crossover scale
$L_{\mathrm{c}}$ diverges  when
 $
K\to1
$ from above.
Thus the system remains in the weak-coupling regime and the
usual perturbative result proportional to $J_0^2$ can be recovered.

\end{document}

%% file: WGL.bbl
%